 \documentstyle[12pt]{article}
\def\prb{Phys. Rev. B }
\def\prl{Phys. Rev. Lett. }

\begin{document}
\baselineskip .3in
\pagestyle{empty}
\sloppy
\newpage
\pagestyle{plain}

\title{Electronic transport in a randomly amplifying and absorbing chain}

\author{Asok K. Sen$^*$ \\
{\it LTP Division, Saha Institute of Nuclear Physics}\\
{\it 1/AF, Bidhannagar, Calcutta 700 064, India}}

\date{\today}
\maketitle

\begin{abstract}
We study localization properties of a one-dimensional disordered system
characterized by a random non-hermitean hamiltonian where both the
randomness and the non-hermiticity arises in the local site-potential;
its real part being ordered (fixed), and a random imaginary part implying
the presence of either a random absorption or amplification at each site.
The transmittance (forward scattering) decays exponentially in either case.
In contrast to the disorder in the real part of the potential (Anderson
localization), the transmittance with the disordered imaginary part may
decay slower than that in the case of ordered imaginary part.
\end{abstract}



$^*${\it e-mail address: asok@cmp.saha.ernet.in}

\newpage
The study of the spectra of systems with non-hermitean hamiltonians and of
the interference of waves multiply scattered from such a system of scatterers 
have of late become very fasionable. The physical reason for such a 
description lies in the fact that the scattering in any real medium is
never perfectly elastic and that in many cases the deviation from perfectly
elastic scattering may be described, for example, by absorption through
other inelastic channels or by amplification due to enhancement of the 
wave-amplitude (e.g., population inversion in an active medium) of incident 
particles or waves.  We are interested in the class of non-hermitean
hamiltonians in which the non-hermiticity is in the local part (typically 
in one-body potentials) \cite{john, expt, weav, kum, frei1, gj, mpl1,pass, 
zek, zhn, been, frei2, misi, yose, frei3}.  It is well known that an
imaginary term in the local part of the hamiltonian behaves like a source
or a sink (depending on the sign).  It may be noted that this type of
complex potentials, called {\it optical potentials}, have been extensively 
studied for isolated atoms in nuclear physics.  For obvious reasons, a
medium having scattering potentials with positive imaginary part $\eta$ 
(sink) at each site is called an {\it absorbing medium} and a medium with
negative $\eta$ (source) {it amplifying medium}.

      In a disordered chain with random but real-valued site-potentials,
almost all the states are exponentially localized and hence an incident
wave ($\sim e^{ikx}$) propagating in the positive {\it x-}direction is
completely backscattered due to the well-known localization effects 
\cite{ramps}.  While a purely ordered chain with fixed absorbing
site-potentials (sinks for particles) leads to an exponential decay of
the transmittance (forward-scattering), one naively expects that the
transmittance would increase indefitely if each of the fixed imaginary 
site-potentials is amplifying (source of particles).  Interestingly, it
was shown by the author \cite{mpl1} both analytically and numerically
that asymptotically (in the large length limit) the transmittance decays
exponentially in both the cases  and that the asymptotic decay constants
are identical for an absorbing and an amplifying chain with the same
magnitude ($|\eta|$) for the strength of the non-hermitean term.  This
somewhat surprising duality between the amplifying and the absorbing
(ordered) cases was confirmed later by Paasschens {\it et al} \cite{pass} 
for a classical Helmholtz equation describing propagation of radiation
(light) through a medium with a complex dielectric constant.  While the
above duality was originally obtained \cite{mpl1} for a tight binding
hamiltonian, recently we \cite{zek} observed the same generic behavior
for an ordered Schrodinger hamiltonian as well.  Generically, the
transmittance decays monotonically with length for an absorbing chain,
while it increases in an oscillatory fashion for an amplifying chain
upto a length determined by $|\eta|$, beyond which the transmittance
decays exponentially.  The study of disorder in all the works
considered so far has been constrained to the real part of the (local)
potential (dielectric constant, in the classical case).  In this work,
we generalize over our work in \cite{mpl1} and consider the effects of
randomness in the amplification/ absorption ({\it imaginary disorder}
at each site.

We consider a quantum chain of $N$ lattice points (lattice constant unity),
represented by the standard single band, tight binding equation:
\begin{equation}
      (E - \epsilon_n) c_n = V(c_{n-1} + c_{n+1}).
\end{equation}
To calculate transport, we consider a {\it open quantum system} which 
consists of the above chain coupled to the external world (two reservoirs
at a very slightly different electrochemical potentials) with two
identical semi-infinite perfect leads on either side.  Here $E$ is the
fermionic energy, $V$ is the constant nearest neighbor hopping term which is
the same in both the leads and the sample, $\epsilon_n$ is the site-energy,
and $c_n$ is the site amplitude at the $n-$th site.  Without any loss of
generality, we choose $\epsilon_n = 0$ in the leads and $V=1$ to set the
energy scale.  Inside the sample, we choose $\epsilon_n = \epsilon_r +
i\epsilon_i$ where both the real and the imaginary parts could be random,
and $i=\sqrt{-1}$.  For the purpose of this work, there is no disorder in
the real part and we take for simplicity $\epsilon_r = 0$.  The imaginary 
part $\epsilon_i$ has the form $[\eta - W_i/2 , \eta + W_i/2]$, where the
constant part $\eta$ may be either positive or negative, or zero and $W_i$ 
is the width of the uniform random distribution in $\epsilon_i$.  The
complex transmission amplitude in the ordered case ($W_i = 0$) was
calculated in \cite{mpl1} to be
\begin{equation}
t_A={(e^{ik_s}e^{-\gamma}-e^{-ik_s}e^{\gamma})(e^{ik}-e^{-ik})e^{-ik(L+2)}
\over de^{-ik_sL}e^{\gamma L} - ce^{ik_sL}e^{-\gamma L}},
\end{equation}
\noindent where
\begin{equation}
   c=(e^{ik_s-ik}e^{-\gamma} - 1)^2, d=(e^{-ik_s-ik}e^{\gamma} - 1)^2,
\end{equation}
\noindent and the decay length $1/|\gamma|=l_a$ and the wave-vector $k_s$
are given by
\begin{equation}
	E = 2{\rm cos} k = (e^{\gamma} + e^{-\gamma}){\rm cos} k_s~,
\end{equation}
and
\begin{equation}
	 \eta = (e^{\gamma} - e^{-\gamma}){\rm sin} k_s~.
\end{equation}
\noindent The transmittance or the two-probe conductance $T=g_2=|t_A|^2$
obtained from the Eq.(2) is found to decay monotonically (exponentially)
towards zero for a set of absorbers ($\eta > 0$).  But, for a set of
amplifiers ($\eta < 0$), $g_2$ increases first to a high value but
eventually (for large $L$) decays as $t_A \sim e^{-|\gamma|L}$.

Disorder in the real part ($\epsilon_r$) in 1D is well-known to give
rise to an exponential decay \cite{ramps}.  Let us consider the case of
a disorder in the imaginary term of the potential with $\eta = 0$, but
$W_i \ne 0$.  It may be noted that in this case (for a long enough chain),
about half of the sites act as absorbers ($\eta > 0$) and about half as
amplifiers ($\eta < 0$).  Then as discussed above, the net contribution
to the transmittance from all the sites would essentially be decaying
with a superposition of various decay constants.  The modes with the 
fastest decay rates will possibly dominate the net transmittance.  When 
$\eta \ne 0$, the distribution of all the decay constants would be
asymmetric, and the net decay constant is expected to be somewhat
different from the $\eta = 0$ case.  In the Fig.~1, we have shown the
various cases with $\eta = 0.01$ and $W_i = 0.3$.  We find that the
pure imaginary case ($\eta = 0.01$, $W_i = 0$) gives rise to a decay
length $1/2\gamma = 100$ as obtained from the equations above.  For the
symmetrically disordered case ($\eta = 0$, $W_i = 0.3$), the decay 
length is about 440.  Clearly the latter decay length in the disordered
case is much larger than the same for the pure imaginary case (somewhat
counter-intuitive).  Finally for the asymmetrically disordered case
($\eta = 0.01$, $W_i = 0.3$), then decay length is about 120 which is
in between the two extreme cases.

In the Fig.~2, we have considered another situation with the same
$\eta = 0.01$ but a different disorder $W_i = 0.7$.  For the symmetrical
disorder case ($\eta = 0$, $W_i = 0.7$), the decay length is about 85
which is smaller than that in that in the pure imaginary case.  Indeed,
this is what one normally expects to be the role played by disorder
(in the real part).  Further, in contrast to that of Fig.~1, the
transmittance decays faster in the asymmetrically disordered cases
($\eta = \pm 0.01$, $W_i = 0.7$) than in the symmetrically disordered
case ($\eta = 0$, $W_i = 0.7$), the decay length in the former being 
about 80.

To summarise, we have studied the transmittance through a 1D chain
with randomly amplifying and/ or absorbing site-potentials at each site.
We find that in contrast to the real disordered potential case, the
decay of transmittance (exponential localization) could be slower
in the complex disordered case (compared to the complex ordered case).
This, in particular, implies that the scattering from the disorder
of this type (i.e., imaginary) may never be incoherent.  Thus, there is
no cut-off length scale for a crossover from a localized to a diffusive
behaviour even when the chain consists of both absorbing as well as
amplifying local potentials in a random fashion.

\smallskip

{\bf Acknowledgements}

The author would like to thank the organisers of the CM Days 97, and
the warm hospitality of the Department of Physics, Vishwa-Bharati
University, Santiniketan, during the progress of the workshop.

\newpage

{\bf Figure Captions:}

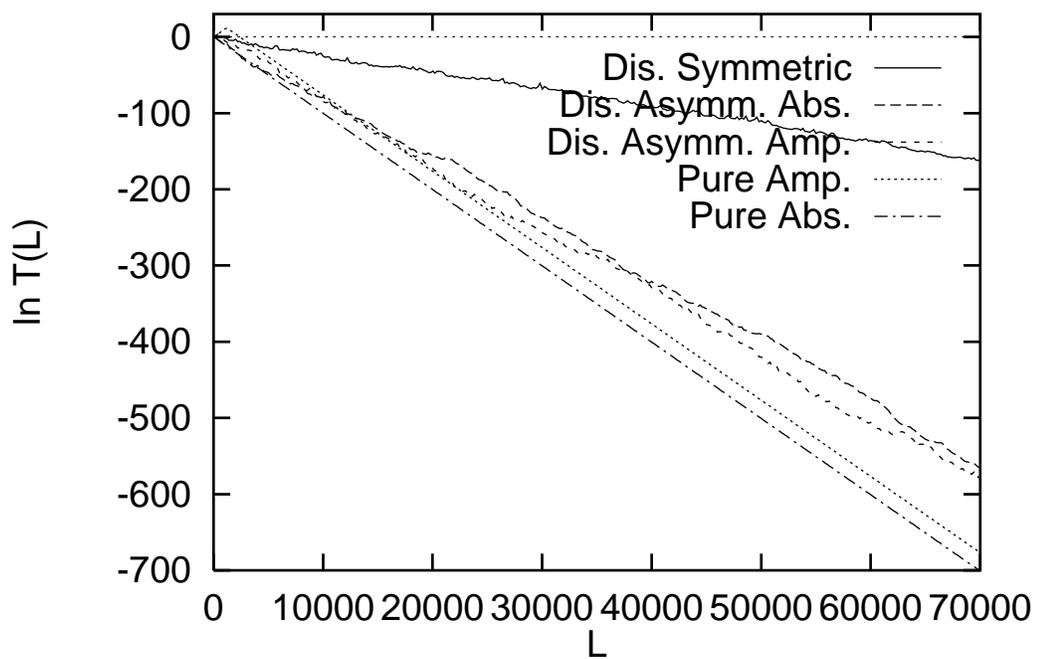
{\bf Fig.1} The variation of the logarithmic transmittance as a function
of $L$ in units of the lattice constant for various combinations of
$\eta = 0$, $W_i = 0.3$.  There is no disorder in the real part of the
site-energy.  The pure absorbing/ amplifying case means that $\eta =
\pm 0.01$, $W_i = 0$; the symmetric disordered case means that $\eta = 0$, 
$W_i = 0.3$; and the asymmetric disordered absorbing/ amplifying case imply
$\eta = \pm 0.01$, $W_i = 0.3$.  Note that the transmittance decays faster
in the pure cases than in the disordered ones.

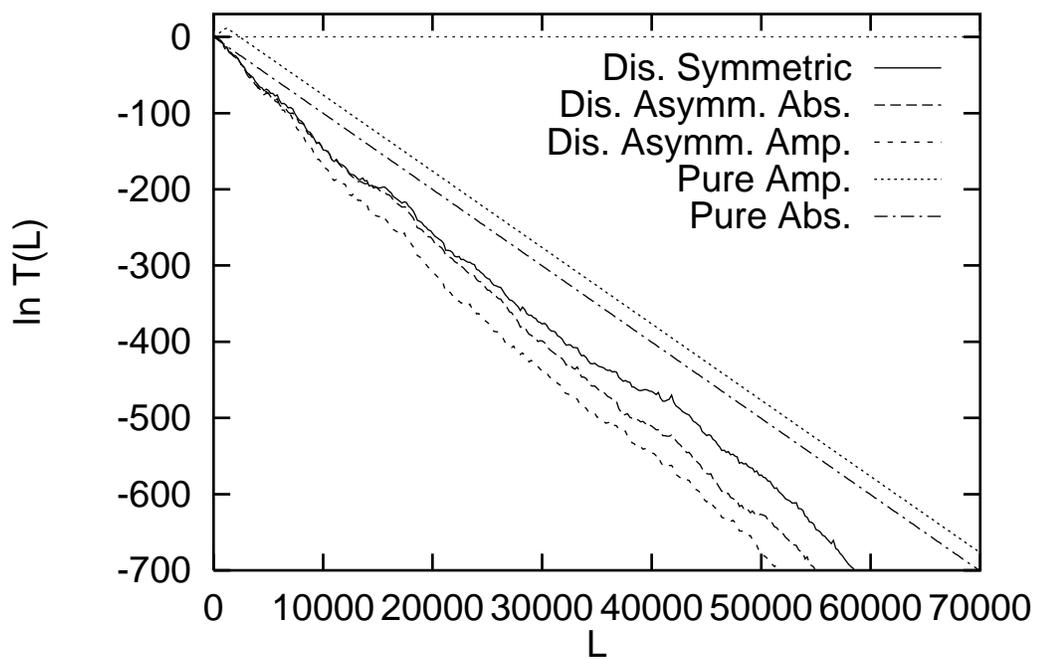
{\bf Fig.2} The same as in Fig.~1, but for different combinations of $\eta
= \pm 0.01$, $W_i = 0.7$.  Again, there is no disorder in the real part of
the site-energy.  For these parameters, the transmittance decays faster
for the (imaginary) disordered cases than in the pure ones.


\begin{thebibliography}{9}

\bibitem{john} S. John, \prl {\bf 53}, 2169 (1984).

\bibitem{expt} A. Z. Genack, \prl {\bf 58}, 2043 (1986); A. Z. Genack and
Garcia, \prl {\bf 66}, 2064 (1991); N. M. Lawandy, R.M. Balachandran, A.S.L.
Gomes, and E. Sauvin, Nature {\bf 368}, 436 (1994); D. S. Wiersma, M. P. van
Albada and A. Lagendijk, \prl {\bf 75}, 1739 (1995).

\bibitem{weav}R. L. Weaver, \prb {\bf 47}, 1077 (1993).

\bibitem{kum} A. Rubio and N. Kumar, Phys. Rev. B{\bf 47}, 2420 (1993); P.
Pradhan and N. Kumar, Phys. Rev. B{\bf 50}, 9644 (1994).

\bibitem{frei1} V. Freilikher, M. Pustilnik, and I. Yurkevich, \prl
{\bf 73}, 810 (1994).

\bibitem{gj} A. Kar Gupta, A.M. Jayannavar, Phys. Rev. B{\bf 52}, 4156
(1995).

\bibitem{mpl1} A.K. Sen, Mod. Phys, Lett. B {\bf 10} 125 (1996); A.K. Sen,
ICTP preprint no. IC/95/391 (1995).

\bibitem{pass} J.C.J. Paasschens, T. Sh. Misirpashaev and C.W.J. Beenakker,
\prb {\bf 54}, 11887 (1996).

\bibitem{zek} N. Zekri, H. Bahlouli and A.K. Sen, ICTP preprint no. IC/97/131
(1997); a slightly modified version appeared in J. Phys.: Condens. Matter
{\bf 10}, 2405 (1998).

\bibitem{zhn} Z. Q. Zhang, \prb {\bf 52}, 7960 (1995).

\bibitem{been} C.W.J. Beenakker, J.C.J. Paasschens and P.W. Brouwer, \prl
{\bf 76}, 1368 (1996).

\bibitem{frei2} V. Freilikher, M. Pustilnik, and I. Yurkevich, preprint
cond-mat/9605090.

\bibitem{misi} T. Sh. Misirpashaev, J.C.J. Paasschens and C.W.J. Beenakker,
Physica A {\bf 236}, 189 (1997).

\bibitem{yose} M. Yosefin, Europhys. Lett. {\bf 25}, 675 (1994).

\bibitem{frei3} V. Freilikher and M. Pustilnik, \prb {\bf 55}, 653 (1997).

\bibitem{ramps} P.A. Lee and T.V. Ramakrishnan, Rev. Mod. Phys. {\bf 57},
287 (1985). See also {\it Scattering and Localization of Waves in Random
Media}, ed. P. Sheng (World Scientific, Singapore, 1990).
 
\bibitem{physa} A.K. Sen, ICTP preprint no. IC/97/130 (1997); Physica A, in
press (1998).

\end{thebibliography}
\end{document}